\documentclass{elsarticle}
\usepackage{amssymb,amsmath,amsthm,leftidx}
\theoremstyle{definition}
\newtheorem{defn}{Definition}[section]
\usepackage{bm}
\usepackage{calc}
\usepackage{tikz}
\usetikzlibrary{trees}
\usetikzlibrary{decorations.pathmorphing}
\usetikzlibrary{decorations.markings}
\newcommand{\SummerTime}{\texttt{SummerTime}}
\newcommand{\Mathematica}{\textit{Mathematica}}

\newcommand{\Sum}{\mathop{\Sigma}\nolimits}
\newcommand{\SumDown}{\Sum_{<} }
\newcommand{\SumUp}{\Sum_{\geqslant} }

\newcommand{\sgn}{\mathop{\mathrm{sgn}}}
\newcommand{\mm}[1]{\mbox{\texttt{\textbf{#1}}}}
\newcommand{\mms}[1]{\mbox{\tiny\texttt{#1}}}

\allowdisplaybreaks
\begin{document}
\newcommand{\sBox}[2]{\newsavebox{#1}\savebox{#1}{\raisebox{0.4ex-0.5\height}{\begin{tikzpicture}#2\end{tikzpicture}}}}

\tikzset{
	every child/.style={very thick},
	edgeFrom/.style={very thick,midarrow},
	edgeTo/.style={very thick,midarrow},
	level/.style={level distance=14mm,
		sibling distance=4ex},
	%	growth parent anchor/.style={},
	every node/.style={thick,rectangle, rounded corners,draw,inner sep=2pt}
}

\sBox{\grOne}{
	\node {$0$} [grow=right]
	child [-stealth,level distance=10mm] {node {$k_3+1$}
		child [stealth-] {node {$k_2+1$}		child [-stealth] {node {$k_1+2$}}}};
}

\sBox{\grTree}{
	\node {$0$} [grow=right]	child [stealth-] {node {$k_1$} child [stealth-] {node {$k_2$}} child [stealth-] {node {$k_3$}	}};
}

\sBox{\grTwo}{
	\node {$0$} [grow=right]	child [-stealth,level distance=10mm] {node {$k_3+1$}}	child [stealth-,level distance=10mm] {node {$k_2$}		child [-stealth,level distance=15mm] {node {$k_1+1$}}};
}
\sBox{\grThree}{
	\node {$0$} [grow=right]	child [-stealth,level distance=10mm] {node {$k_2+1$} child [-stealth] {node {$k_1+2$}} child [-stealth] {node {$k_3+1$}	}};
}

\sBox{\grFour}{
	\node {$0$} [grow=right]	
	child [-stealth,level distance=10mm] {node {$k_3+1$}}	
	child [stealth-,level distance=10mm] {node {$k_2$}}		
	child [-stealth,level distance=10mm] {node {$k_1+1$}};
}

\sBox{\grFive}{
	\node {$0$} [grow=right]
	child [-stealth,level distance=12mm] {node {$k_3+1$}}
	child [stealth-,level distance=12mm] {node {$k_1$}
		child [stealth-] {node {$k_2-1$}}};
}
\sBox{\grSix}{
	\node {$0$} [grow=right]	child [stealth-,level distance=12mm] {node {$-k_2-1$} child [stealth-,level distance=17mm] {node {$-k_1-2$}} child [stealth-,level distance=17mm] {node {$-k_3-1$}	}};
}

\sBox{\grSeven}{
	\node {$0$} [grow=right]	
	child [stealth-,level distance=14mm] {node {$-k_3-1$}}	
	child [stealth-,level distance=14mm] {node {$k_2$}}		
	child [stealth-,level distance=14mm] {node {$-k_1-1$}};
}

\sBox{\grEight}{
	\node {$0$} [grow=right]	child [stealth-] {node {$-k_3-1$}}	child [stealth-] {node {$k_1$}		child [stealth-] {node {$k_2-1$}}};
}
\begin{frontmatter}

\title{Introducing \SummerTime: a package for high-precision computation of sums appearing in DRA method %\tnoteref{mytitlenote}
}
%\tnotetext[mytitlenote]{title footnote}


\author[binp]{Roman N. Lee}
\ead{r.n.lee@inp.nsk.su}

\author[binp,nsu]{Kirill T. Mingulov}
\ead{melkogotto@gmail.com}

\address[binp]{Budker Institute of Nuclear Physics, 630090, Novosibirsk}
\address[nsu]{Novosibirsk State University, 630090, Novosibirsk}
\begin{abstract}
We introduce the \Mathematica\ package \SummerTime\ for arbitrary-precision computation of sums appearing in the results of DRA method. So far these results include the following families of the integrals: 3-loop onshell massless vertices, 3-loop onshell mass operator type integrals, 4-loop QED-type tadpoles, 4-loop massless propagators \cite{LeeSmSm2010,LeeSmi2010,LeeSmSm2011, Lee2011e}. The package can be used for high-precision numerical computation of the expansion coefficients of the integrals from the above families around arbitrary space-time dimension. In addition, this package can also be used for calculation of multiple zeta values, harmonic polylogarithms and other transcendental numbers expressed in terms of nested sums with factorized summand.
\end{abstract}

%\begin{keyword}
%Multiloop calculations, DRA method, hypergeometric sums
%\end{keyword}

\end{frontmatter}

\section{Introduction}

%\begin{tikzpicture}[>=stealth, every node/.style={circle, draw, minimum size=0.75cm}]
%\graph [tree layout, grow=down, fresh nodes, level distance=0.5in, sibling distance=0.5in]
%{
%	4 -> { 
%		3 -> { 1 -> { 5, " " }, 2,2 },
%		3 -> { 1, 2, 2 },
%		3 -> { 1, 2, 2 }
%	} 
%};
%\end{tikzpicture}
Multiloop corrections within the Standard model and in general quantum field theory are of great interest now. In particular, this interest is connected with continuing search of New Physics both in collider experiments, such as LHC, and in high-precision spectroscopy measurement. 

Probably, the most effective approach to the multiloop calculations is the differential equations technique. It was introduced long ago \cite{Kotikov1991,Kotikov1991a,Kotikov1991b,Remiddi1997} and has an impressive record of achievements. However, the differential equations do not work for one-scale integrals as their dependence on the scale is totally determined by dimensional considerations. On the other hand, the one-scale integrals are ubiquitous: they appear both in physical observables, like lepton anomalous magnetic moments or asymptotic $R(s)$ ratio, and as boundary conditions in differential equations for multiloop integrals.

Some time ago the DRA method of the calculation of the multiloop integrals has been introduced in Ref. \cite{Lee2010}. This method is based on using the dimensional recurrence relations and analytical properties of the integrals as functions of the space-time dimensionality $d$. In a short time, the DRA method has been successfully applied to a wide range of physically interesting one-scale families of integrals \cite{LeeSmSm2010,LeeSmi2010,LeeTer2010,LeeSmSm2010a,LeeSmSm2011,Lee2011e,LeeSmirnov2012,Lee2012a,LeeMarquardSmirnovSmirnovSteinhauser2013}. 

The results of the DRA method are exact in $d$ and have a form of nested sums with factorized dependence of the summand on the summation variables. Meanwhile, the applications require series expansion of the integrals in $d$ near integer point, usually near $d=4$. The most effective approach to obtaining analytical form of this expansion in terms of conventional transcendental numbers proved to be the one based on high-precision calculation of the expansion coefficients and subsequent use of the \textrm{PSLQ} algorithm. Many analytical results for the $\epsilon$-expansions of the master integrals obtained in this way have been already published in the papers \cite{LeeSmSm2010,LeeSmi2010,LeeTer2010,LeeSmSm2010a,LeeSmSm2011,Lee2011e,LeeSmirnov2012,Lee2012a,LeeMarquardSmirnovSmirnovSteinhauser2013} near $d=4$ (in Ref. \cite{LeeTer2010} also near $d=3$). However, the results of DRA method contain much more information than the published expansion coefficients. In fact, they are quite analogous to the representation in terms of hypergeometric functions, and totally determine expansion coefficients around any value of $d$. The only problem with these results is that, similar to the expansion of the hypergeometric functions, it may be quite difficult to express these coefficients in terms of conventional transcendental numbers, like multiple zeta values. Meanwhile, from time to time, there appears a necessity to calculate expansions of the master integrals either in different dimensionality or up to a higher order in $\epsilon$.

The main goal of the present paper is to introduce a \Mathematica{} package \SummerTime{} for arbitrary-precision calculation of the expansion coefficients of DRA results around arbitrary value of $d$. This package gives high-energy community the full access to the results of the DRA method. In addition, the package contains procedures for high-precision evaluation of the multiple-zeta values and harmonic polylogarithms. 

\section{Solution of dimensional recurrence relation}

Dimensional recurrence relations seem to appear for the first time in Ref. \cite{DerkachovHonkonenPismak1990} and then rediscovered independently and systematically used in Ref. \cite{Tarasov1996}. 
%In order to derive the dimensional recurrence relation one has to use standard IBP reduction procedure. Within this approach, one considers a set of integrals labeled by integer multiindex $(n_1,\ldots,n_N)$:
%\begin{equation}
%J^{(d)}(n_1,\ldots, n_N) = \int \frac{\dd^dl_1\ldots \dd^dl_L}{\pi^{Ld/2}D_1^{n_1}\ldots D_N^{n_N}}\,, 
%\end{equation}
%where $D_1\ldots D_N$ is a complete set of linear function of $N=L(L+1)/2+L E$ scalar products 
%\begin{equation}
%l_i\cdot q_j,\quad  1\leq i\leq j\leq L+E\,.
%\end{equation}
%Here $q_1,\ldots, q_L=l_1,\ldots,l_L$ and $q_{L+1},\ldots, q_{L+E}=p_1,\ldots, p_E$, where $p_1,\ldots, p_E$ are independent external moments. The IBP identities connect integrals with shifted indices and allow one to reduce any integral to a finite set of the master integrals. They also allow one to obtain the dimensional recurrence relations for the master integrals. 
For the column-vector $\bm{J}$ of the master integrals of a given sector (i.e., the integrals with a given set of denominators) the dimensional recurrence relation has the form
\begin{equation}\label{eq:DRR}
\bm J(\nu+1)=\mathbb{C}(\nu)\bm J(\nu)+\bm  R(\nu)\,,
\end{equation}
where inhomogeneous term $\bm R(\nu)$ contains master integrals from the subsectors, $\nu=d/2$. Let $\mathbb{S}(\nu)$ be a revertible matrix, satisfying the equation
\begin{equation}
	\mathbb{S}(\nu)=\mathbb{S}(\nu+1)\mathbb{C}(\nu).
\end{equation}
Then the general solution of Eq. \eqref{eq:DRR} can be written as follows:
\begin{equation}\label{eq:solution}
\bm J(\nu) = \mathbb{S}^{-1}(\nu)\Sum \mathbb{S}(\nu)\mathbb{C}^{-1}(\nu)\bm  R(\nu).
\end{equation}
Here $\Sum$ is the indefinite sum symbol with the property
\begin{equation}
	(\Sum F)(\nu+1)-(\Sum F)(\nu)=f(\nu).
\end{equation}
If the function $F(\nu)$ decreases faster than $1/\nu$ when $\nu\to +\infty$  and/or $\nu\to -\infty$, one can write
\begin{equation}
\Sum F (\nu)=\SumUp F (\nu)+\omega(z)\stackrel{\text{def}}{=}-\sum_{n=0}^{+\infty}F(\nu+n)+\omega(z)\label{eq:SumUp}
\end{equation}
and/or
\begin{equation}
\Sum F (\nu)=\SumDown F (\nu)+\omega(z)\stackrel{\text{def}}{=}\sum_{n=-\infty}^{-1}F(\nu+n)+\omega(z)\label{eq:SumDown},
\end{equation}
where $\omega(z)=\omega(\exp(2\pi i \nu))$ is arbitrary periodic function. The key feature of the DRA method is the determination of the function $\omega(z)$ from the analytical properties on the integrals as functions of $\nu$.

Assuming that simpler master integrals in $\bm R (\nu)$ are also represented in a form \eqref{eq:solution}, we naturally come to the problem of calculation of the nested sums of the form
\begin{equation}
\mathbb{J}=
\mathbb{S}_K^{-1}\Sum_{\tau_K} \mathbb{S}_K\ldots \mathbb{S}_2^{-1}\Sum_{\tau_2} \mathbb{S}_2\mathbb{T}_{21}\mathbb{S}_1^{-1}\Sum_{\tau_1} \mathbb{S}_1\mathbb{T}_{10}\mathbb{S}_0^{-1}\,.
\end{equation}
Here $\mathbb T_{i+1,i}(\nu)$ are rational rectangular matrices, $\tau_i\in\{\geqslant,<\}$, and $\mathbb{S}_i(\nu)$ is a hypergeometric matrix term, i.e. $\mathbb{S}_i^{-1}(\nu)\mathbb{S}_i(\nu+1)$ is a rational matrix. Explicitly, we have
\begin{multline}
\mathbb{J}(\nu)=
\sum_{n_i\tau_i0}\mathbb{S}_K^{-1}(\nu)\mathbb{S}_K(\nu+n_K)\ldots 
(\mathbb{S}_2\mathbb{T}_{21}
\mathbb{S}_1^{-1})(\nu+n_K+\ldots+n_2)\\
\times
 (\mathbb{S}_1\mathbb{T}_{10}\mathbb{S}_0^{-1})(\nu+n_K+\ldots+n_1)\,.
\end{multline}
Here $n_i\tau_i0$ denotes $n_i<0$ or $n_i\geqslant0$ depending on whether $\tau_i$ is equal to `$<$' or  `$\geqslant$'.
Passing to the variables $k_i=n_i+n_{i+1}+\ldots +n_K$, we obtain
\begin{multline}\label{eq:matrix_sum}
\mathbb{J}(\nu)=
\sum_{k_i\tau_ik_{i+1}}\mathbb{S}_K^{-1}(\nu)\mathbb{S}_K(\nu+k_K)\ldots 
(\mathbb{S}_2\mathbb{T}_{21}
\mathbb{S}_1^{-1})(\nu+k_2)(\mathbb{S}_1\mathbb{T}_{10}\mathbb{S}_0^{-1})(\nu+k_1)\,.
\end{multline}

In many physically interesting families of integrals each sector contains at most one master integral. Then all matrices are replaced by scalars and we come to the problem of evaluating the sums of the following form
\begin{equation}\label{eq:scalar_sum}
	J(\nu)=\sum_{k_i\tau_ik_{i+1}}F_K(\nu+k_K)\ldots F_1(\nu+k_1)\,,
\end{equation}
where each $F_i$ is a hypergeometric term, i.e. the ratio  $F_i(\nu+1)/F_i(\nu)$ is a rational function, and $k_{K+1}\equiv0$. In particular such are all the families of integrals calculated within DRA method \cite{LeeSmSm2010,LeeSmi2010,LeeTer2010,LeeSmSm2010a,LeeSmSm2011,Lee2011e,LeeSmirnov2012,Lee2012a,LeeMarquardSmirnovSmirnovSteinhauser2013} except for Ref. \cite{LeeSmirnov2012} where the result for multimaster integral was obtained.  The present version of \SummerTime{} package is restricted to the calculation of the scalar sums \eqref{eq:scalar_sum}.

\section{Tree sums}
Remarkable property of the sums \eqref{eq:matrix_sum} and \eqref{eq:scalar_sum} is that the dependence on the summation variables in the summand is factorized. If the limits of the sums were decoupled, the sum would be a product of one-fold sums. Nevertheless, even with the limits given by inequalities  $k_i\tau_ik_{i+1}$, the sums in Eqs. \eqref{eq:matrix_sum} and \eqref{eq:scalar_sum} can be organized without nested loops as we explain below. But first we introduce the notion of tree sums and explain how sums of the form \eqref{eq:matrix_sum} or  \eqref{eq:scalar_sum} can be rewritten in terms of those.

Let us define the following correspondence between a certain vertex-labeled directed graph and the solution of a set of inequalities. Namely, let us consider a directed graph with nodes marked by some expressions. Then the corresponding set of inequalities contains inequality $m\geqslant n$ for each edge $m\to n$. E.g., the set of integer triples $(k_1,k_2,k_3)$ satisfying inequalities $(k_3<0)\&(k_2\geqslant k_3)\&(k_1<k_2)$ corresponds to the graph \usebox{\grOne}. All sums appearing in the DRA method can be labeled by path graphs. However, convergence rate analysis is easier for the sums which summation limits determined by the directed rooted tree graphs. Any path graph can be reduced to a set of directed tree graphs. E.g.,
\begin{multline}
\usebox{\grOne}=\usebox{\grTwo}+\usebox{\grThree}\\
=\usebox{\grFour}+\usebox{\grFive}+\usebox{\grThree}
\\
=\usebox{\grSeven}+\usebox{\grEight}+\usebox{\grSix}
\end{multline}
which corresponds to the identity
\begin{multline}
\label{eq:example:reduction_to_tree_sums}
\sum_{\substack{
k_3<0\\k_2\geqslant k_3\\k_1<k_2
}} F(k_3,k_2,k_1)
=
\sum_{\substack{0\leqslant n_3\\0\leqslant n_2\\0\leqslant n_1}} F(-n_3-1,n_2,-n_1-1)\\
+\sum_{\substack{0\leqslant n_3\\0\leqslant n_1\\n_1\leqslant n_2}} F(-n_3-1,n_2+1,n_1)
+\sum_{\substack{0\leqslant n_2\\n_2\leqslant n_1\\n_2\leqslant n_3}} F(-n_3-1,-n_2-1,-n_1-2)\,.
\end{multline}
Note that if the summand on the left-hand side of Eq. \eqref{eq:example:reduction_to_tree_sums} was a hypergeometric term with factorized dependence on the summation variables, the same is also true for the summands on right-hand side of the equation. The factorized form of the summand is a very important property which allows one to organize the calculation of the sums without nested loops. 

In what follows we will use term `tree sum' for the sums with limits determined by the directed rooted tree and with factorized dependence of the summand on the summation variables. It is very instructive to use the following alternative recursive definition of tree sums:
\begin{defn}\label{def:tree_sum}
A \emph{tree sum} is
\begin{itemize}
\item \textbf{Case 1}: Any one-fold sum $ T(k)=\sum_{n=k}^{\infty} a(n) $
\item \textbf{Case 2}: Any sum of the form $ T(k)=\sum_{n=k}^{\infty} a(n) \prod_{i=1}^j T_i(n)$, where $ T_i(n)$ are tree sums ($j$ is a natural number).
\end{itemize} 
\end{defn}

\subsection{Computation of tree sums}

The numerical estimate of the tree sum can be obtained by replacing in the above definition the upper limit $\infty$ by some sufficiently large number $N$.  A naive prescription would give the computational complexity $O(N^K)$ of such an estimate, where $K$ is a `nestedness' of the sum (the number of summation variables). In order to organize the computation much more effectively, we use recursive nature of the definition \ref{def:tree_sum}. We first note that the evaluation of the list of one-fold sums $\{T(0),T(1),\ldots,T(N)\}$ can be done in one path with the computational complexity $O(N)$. Then, assuming that we already have such lists for each $T_i$, we can evaluate  $T(k)$ from Case 2 of definition \ref{def:tree_sum} also in $O(N)$. Therefore, the overall computational complexity of obtaining the estimate is $O(N K)$. We stress the difference between the computational complexities of the sums with factorized and nonfactorized summand.

\paragraph{Convergence acceleration}
Once we eliminated the exponential dependence on $K$, we have to examine also the dependence of the complexity on the required precision $P$ ($P$ is a number of decimal digits). For exponentially converging sums (in what follows we call them \emph{G-sums}) we have $N\sim P$ and the overall computational complexity is\footnote{We do not take into account the penalty of dealing with the $P$-digit numbers, which makes the complexity likely to be $O(P^2 K)$.}  $O(P K)$. However, for power-like behavior of the summand (we call such sums the \emph{H-sums}) the number of terms in the naive computation depends exponentially on $P$. Therefore, we need to devise a suitable algorithm for convergence acceleration. 

Let us consider a one-fold sum 
\begin{equation}
T=\sum_{n=0}^{\infty}a(n) \,,\quad a(n) \sim \frac1{n^{\alpha+1}}\text{ at large }n\,,\ \alpha>0\,.
\end{equation}
Then in a naive prescription the required number of terms is $N\sim 10^{P/\alpha}$. Fortunately, it appears that we can use the  convergence acceleration described in Ref. \cite{Broadhurst:1996}. The general idea of the convergence acceleration is to replace the original sum with the modified one converging to the same value but with faster convergence rate. Let the transformed sum have the form 
\begin{equation}
T_1=\sum_{n=0}^{\infty} a_1(n)
\end{equation} and 
\begin{equation}
a_1(n) = (1-f(n))a(n)+f(n+1)a(n+1)\,,
\end{equation}
where $f(n)$ is some function to be fixed and $a(-1)=0$ by definition. Then the partial sum $T_1^{(N)}$ is expressed via $T^{(N)}$ as 
\begin{equation}
T_1^{(N)}=\sum_{n=0}^{N} a_1(n)=T^{(N)}-f(0)a(0)+f(N+1)a(N+1).
\end{equation}
The partial sums $T_1^{(N)}$ an $T^{(N)}$  both have the same limit when \begin{equation}
f(0)=0\,,\quad f(n)a(n)\stackrel{n\to\infty}\longrightarrow 0\,.
\end{equation} 
The convergence rate is boosted when
\begin{equation}
 a_1(n)/a(n)\stackrel{n\to\infty}\sim 1-f(n)+f(n+1)(1+1/n)^{-\alpha-1} \stackrel{n\to\infty}\longrightarrow 0\,.
\end{equation}
All these conditions are satisfied by $ f(n) = n/\alpha$. Note that if we know $N$ terms of the original sequence $a(n)$, we can reconstruct $N-1$ terms of the transformed sequence.

The leading asymptotics of the transformed term $a_1(n)$ depends on the next-to-leading term of $a(n)$. Let us assume that the asymptotic expansion of $a(n)$ has the form
\begin{equation}
  a(n) \sim \sum_{k=1}^{\infty} \frac{c_{0,k}}{n^{\alpha+k}}.
\end{equation} 
Then the transformed sequence has the expansion
\begin{equation}
  a_1(n) \sim \sum_{k=1}^{\infty} \frac{c_{1,k}}{n^{\alpha+1+k}}\,.
\end{equation} 
This allows one to apply the same acceleration technique recursively. If we have $N$ terms of the sequence, we may apply the above acceleration technique $N-1$ times which gives an estimate $N\lg N\sim P$. Taking into account that computational complexity of the acceleration is $O(N^2)$, we have a rough estimate $O(P^2/\lg^2 P)\lesssim O(P^2)$ for the overall complexity of the power-like decaying sums. Note that the convergence acceleration is compatible with our algorithm of evaluation of the nested tree sums. The only complication comes from the fact that convergence acceleration leads to strong numerical compensations when calculating the terms of the accelerated sequence, therefore the working precision $W$ should be taken higher then the required one, and the empirical rule is $W =\gamma P$, where $\gamma\approx 1.7$. Then, for $K$-fold sum we have $O(K \gamma^K P^2)$ complexity estimate.

\paragraph{Expansion in $\epsilon$}

Tree sums appearing in the DRA method depend on space-time dimensionality $d$. One is usually interested in their Laurent expansion near some point (typically, near $d=4$). For a generic sum, depending on $\epsilon$, the Laurent expansion does not always commute with the summation. However, the $\epsilon$ dependence  of the tree sums appearing in the DRA method is very special: any summation variable $n$ enters the summand only via combination $n+\epsilon$ or $n-\epsilon$. Therefore, the sums converge uniformly in $\epsilon$ which means that the Laurent expansion and the summation commute. Therefore, in order to calculate the former, one can first expand the summand and then make the summation. It is important that our convergence acceleration procedure is linear. Therefore we can apply it to formal power series in exactly the same way as to numbers.

\section{Conventional transcendental numbers and functions as tree sums}

The main purpose of the \SummerTime{} package is to calculate sums which arise in the DRA method. However, it is remarkable that many conventional transcendental numbers and functions can be expressed and effectively evaluated as tree sums. Let us consider the Goncharov polylogarithms defined via 
\begin{equation} 
\label{eq:goncharov}
\mathrm{Li}_{n_1,\ldots,n_k}(x_1,\ldots,x_k)
\quad = \quad 
\sum_{0 < i_1 < i_2< \ldots < i_k }
 \frac{x_1^{i_1}x_2^{i_2}
\ldots x_k^{i_k}}{i_1^{n_1}i_2^{n_2}\ldots i_k^{n_k}}
\end{equation}
To represent the above definition in the form of a tree sum, it suffices to rewrite it as 
\begin{equation} 
\label{eq:goncharov_tree}
\mathrm{Li}_{n_1,\ldots,n_k}(x_1,\ldots,x_k)
\quad = \quad 
\sum_{0 \leqslant i_1\leqslant ... \leqslant i_k }
 \prod_{j=1}^{k}\frac{x_j^{i_j+j}}{(i_j+j)^{n_j}}\,.
\end{equation}
Multiple zeta values
\begin{equation}
\zeta \left(n_k, \ldots,n_1 \right) = 
	\sum_{i_k >  \ldots > i_1 > 0}\prod_{j=1}^k \frac{\left(\sgn n_j\right){}^{i_j}}{i_j^{ \left| n_j \right| }}
\end{equation}
are trivially related to the Goncharov polylogarithms:
\begin{equation}
\zeta \left(n_k, \ldots,n_1 \right) =
\mathrm{Li}_{|n_1|,\ldots,|n_k|}(\sgn n_1,\ldots,\sgn n_k)\,,
\end{equation}
and, therefore can be readily represented in the form of tree sums.
%
%
%see, e.g., Ref. \cite{Maitre:2005}\todo{check reference}. To make the above definition in the form of tree sum, it suffices to rewrite it as 
%\begin{equation}
%	\label{eq:mzv_def_reduced}
%	\sum_{ 0\leqslant n_1 \ldots \leqslant n_k}
%	\prod_{j=1}^{k} \frac{\left(\sgn m_j\right)^{n_j+j}}{\left(n_j+j\right)^{\left|m_{j}\right|}}\,.
%\end{equation}
Harmonic polylogarithms (HPL) are defined by (see, e.g., Refs. \cite{RemiddiVermaseren:HPL, Maitre:HPL})
\begin{equation}
	H(\underbrace{0,\ldots,0}_{n};x)=\frac{(\ln x)^n}{n!}
\end{equation}
and
\begin{equation}
	H(n, n_k, \ldots, n_1; x) = \int_0^x f_n(t) H(n_k, \ldots, n_1; x),
\end{equation} 
where $n \in \{0, \pm 1\}$ and
\begin{equation}
	f_{1}(t)  = \frac{1}{1-t} \,,\quad f_{0}(t)  = \frac{1}{t} \,,
	\quad f_{-1}(t) = \frac{1}{1+t}\,.
\end{equation}
It is standard to use nonzero integer $n$ in the indices of $H$ as a shortcut for sequence of $|n|-1$ zeros appended by $\sgn n$, e.g., $H(3,-2;x)=H(0,0,1,0,-1;x)$. Since trailing zeros can be eliminated with the help of functional relations, see \cite{Maitre:HPL}, we may consider only $H(n_k, \ldots, n_1; x)$, where all $n_1, \ldots, n_k$ are nonzero integers. We have
%\begin{multline}
\begin{equation}
H(n_k,\ldots, n_1; x)=
(\sigma_1 \sigma_2\ldots\sigma_k)\,\mathrm{Li}_{|n_1|,\ldots, |n_k|}(
\sigma_1 \sigma_2,\sigma_2 \sigma_3,\ldots,\sigma_k x
),
\end{equation}
%\end{multline}
where $\sigma_i=\sgn n_i$. Therefore, HPLs also can be expressed via tree sums.

\section{Application example}

So far, the DRA method was applied to the following families of the integrals:
\begin{itemize}
\item To 3-loop onshell massless vertices \cite{LeeSmSm2010};
\item To 3-loop onshell mass operator type integrals \cite{LeeSmi2010};
\item To 4-loop QED-type tadpoles \cite{LeeTer2010};
\item To 4-loop massless propagators \cite{LeeSmSm2011, Lee2011e}.
\item Partly to 4-loop onshell mass operator type integrals \cite{LeeMarquardSmirnovSmirnovSteinhauser2013}.
\end{itemize}

The results of the DRA method for the first four families can be downloaded as \texttt{.zip} files from the package web page. Together with the \SummerTime{} package, they allow one to calculate with arbitrary precision the expansion of the integrals near any value of $d$.

Let us present a minimal example of the application of \SummerTime{} package. We assume that the content of the archive \texttt{M.zip} is put to a working directory of the \Mathematica{} session. Then the following program calculates the expansion of $M_{6,2}$ integral near $d=3$  up to $\epsilon^3$ ($\epsilon=(3-d)/2$) with the precision 1000 digits:
\begin{align*}
\mms{In[1]:= }&\mm{<<SummerTime`}\\
&\mm{SetDirectory[NotebookDirectory[]]}\\
&\mm{TriangleSumsSeries[<<"M62"/.d -> 3-2$\epsilon$,\{$\epsilon$,3\},1000]}\\
\mms{Out[1]:= }&-126.330936333943790321081484798414734532015352412682120...\times\epsilon^{-2}\\
&+449.863309410101651375361946647344871810340547434784147...\times\epsilon^{-1}\\
&+7492.35397292225323538373824972066060688984257832532489815...\\
&-17762.4348499407483506220786909230708553370113393194388...\times\epsilon\\
&-148136.431403928982879168608861673380547613003055965761...\times\epsilon^2\\
&-3048084.17207428077785717945708196781177068038009440576...\times\epsilon^3\\
&+O\left(\epsilon^4\right)
\end{align*}

On the notebook with 2 cores Intel core i7 the run time of the above program was 12 min.

In order to showcase the usability of the package, we have calculated the expansions of all four-loop massless propagator integrals near $d=3$ (the corresponding results for $d=4$ can be found in Refs. \cite{BaikChe2010,LeeSmSm2011}). The results are presented in Appendix. The calculation provides a strong evidence of appearance of new transcendental numbers in addition to conventional ones: multiple zeta values and alternating harmonic sums.

\section{Brief manual of \SummerTime{} package}

The \SummerTime{} package can be downloaded from the site \linebreak \texttt{http://www.inp.nsk.su/{\textasciitilde}lee/programs/SummerTime/} and installed according to the instruction included in a distribution archive.
After the installation, the package can be loaded into \Mathematica\ session by a command:
\begin{verbatim}<< SummerTime`\end{verbatim}
The full list of functions provided by the package can be seen by \texttt{?SummerTime`*}. Here we will consider only the basic ones.

\subsection{Summation specifications}

The \SummerTime{} package supports two types of summation limit specifications.
The first one, called \textit{triangle}, is designed to handle the sums of form \eqref{eq:scalar_sum}.
This summation specification represents a list of variables with $\pm$ signs, e.g.:
\begin{equation}
    \{ \pm n, \pm k, \dotsc \}.
\end{equation}
Signs determine the direction of summation: $\{ \dotsc, \pm n, k, \dotsc \}$ denotes $\pm n \le k < \infty$ and $\{ \dotsc, \pm n, -k, \dotsc \}$ denotes $\mp n \le k < \infty$.

The second specification is used for tree sums, given by definition \ref{def:tree_sum}. It represents a tree, which is written in the following form:
\begin{equation}
    \{ \text{node}, ~ \text{child}_1, ~ \text{child}_2, ~ \dotsc \},
\end{equation}
where children are also trees. If node doesn't have any children, it's simply written as:
\begin{equation}
    \{ \text{node} \}.
\end{equation}
The node itself is a summation variable. The root of a tree is always an integer.
For example, the specification for a tree sum with limits $0 \le n \le k < \infty$ can be written as $\{0,\{n,\{k\}\}\}$.

\subsection{Summation functions}

The \SummerTime\ package defines following eight summation functions:
\begin{verbatim}TreeSum[expr, specs, p]
TreeSumSeries[expr, specs, {e, o}, p]
TriangleSum[expr, specs, p]
TriangleSumSeries[expr, specs, {e, o}, p]
TreeSums[exprs, p]
TreeSumsSeries[exprs, {e, o}, p]
TriangleSums[exprs, p]
TriangleSumsSeries[exprs, {e, o}, p]\end{verbatim}
Here \texttt{expr} denotes a summand, \texttt{specs} is a summation specification (either tree or triangle), \texttt{exprs} is a list of pairs $\{\mathtt{expr}, \mathtt{specs}\}$, \texttt{e} is a small parameter $\epsilon$, \texttt{o} is an expansion order with respect to $\epsilon$, and $p$ is required precision.
Functions names are constructed in the following way: \texttt{Tree} (\texttt{Triangle}) means that function uses tree (triangle) summation specification, \texttt{Series} indicates that the expansion of sum in $\epsilon$ is performed, and \texttt{Sums} instead of \texttt{Sum} designates that several sums are calculated and their total is returned as a result. The advantage of \texttt{Sums} functions is that they perform computations in parallel.

All summation functions have an optional boolean argument\linebreak \texttt{ProgressIndicator}, which denotes whether a progress bar must be shown during computations. Default value is \texttt{True}.

\texttt{Sum} functions also have an optional argument called \texttt{Information}. By this parameter, user can set up a string identifier, which will be shown on top of the progress bar. 
This option can be helpful for a sequence of calls of the summation functions: using it, user can understand which particular sum is currently being calculated.

Parallel functions (\texttt{Sums}) also have an optional boolean argument \texttt{Parallel}, which toggles, whether computation must be parallelized. Default value is \texttt{True}.

\subsection{Other functions}

Currently, the \SummerTime\ package contains also implementation of multiple-zeta values and harmonic and Goncharov polylogarithms. By default these functions are disabled and in order to use them, one needs to set up a value of special variable \texttt{\$Options} before loading the package:
\begin{verbatim}SummerTime`Package`$Options = {"MZV" -> True, "HPL" -> True,
                               "MPL" -> True};
<< SummerTime`\end{verbatim}
Multiple zeta values, harmonic and Goncharov polylogarithms are available via 
\begin{center}
    \begin{tabular}{ll}
        \texttt{MZV[m1, ..., mk]}        & $\zeta\left(m_1, \dotsc, m_k\right)$ \\
        \texttt{HPL[\{m1, ..., mk\}, x]} & $H\left(m_1, \dotsc, m_k; x\right)$  \\
        \texttt{MPL[\{m1,x1\}, ..., \{mk,xk\}, x]} & $\mathrm{Li}_{m_1,\dotsc, m_k}\left(x_1, \dotsc, x_k; x\right)$  \\
    \end{tabular}
\end{center}
%\texttt{HPL} supports so-called ``a''-notation (i.e. when index-vector contains only $\pm1$ and $0$) and ``m''-notation (when zeros before $\pm1$ are omitted). By default, ``m''-notation is used.
Numerical values of these functions  can be obtained via standard \Mathematica\ function \texttt{N}.

\section{Conclusion}

In the present paper we have introduced a \Mathematica{} package \SummerTime{} for calculation of the multiloop integrals obtained within DRA method. The main purpose of the package is an arbitrary-precision calculation of the expansion coefficients of the integrals around any value of $d$. In addition, the package contains convenience tools for the calculation of various transcendental constants and functions: multiple zeta value, harmonic polylogarithms and Goncharov polylogarithms. 

In the future, we plan to extend the package to the computation of the matrix sums. The difficulties connected with the convergence acceleration of the matrix sums seem to be possible to overcome using the approach of Ref. \cite{Tulyakov2011a}.
Such an extension of the package is highly desirable given the families of integrals of actual interest for the present moment. 
\paragraph{Acknowledgments} This work was supported by RFBR grant No. 15-02-07893.

\appendix
\section{Expansion of massless propagator-type master integrals}
Here we present the results for the expansions of the massless four-loop propagator-type master integrals around $d=3$ ($\epsilon=(3-d)/2$). It turns out \cite{Young:2014sia} that these expansions may be of some interest for the applications.

We fix normalization by dividing all expansions with
\begin{equation}
M_{01}(3-2\epsilon)=\frac{\Gamma \left(\frac{1}{2}-\epsilon \right)^5 \Gamma (4 \epsilon -1)}{\Gamma \left(\frac{5}{2}-5 \epsilon \right)}
\end{equation}
All results have been obtained by numerically calculating the expansions of the  results of Ref.~\cite{Lee2011e} and then using \Mathematica{} built-in function \linebreak \texttt{FindIntegerNullVector} with a set including non-alternating and alternating harmonic sums. For some coefficients this approach failed which strongly indicates appearance of new constants in the expansions. We added the unresolved coefficients to the transcendental basis and used the extended basis for subsequent integrals. In the end we have expressed all expansions in terms of standard transcendental numbers
\begin{gather}
\zeta_n=\sum_{i=1}^{\infty}\frac1{i^n}\,,\quad 
a_n=\sum_{k=1}^{\infty}\frac{1}{2^kk^n}\,,
\quad
s_6=\sum_{m=1}^{\infty}\sum_{k=1}^{m}\frac{(-1)^{m+k}}{m^5k}
\end{gather}
and the following ten unresolved coefficients:
\begin{align}
M_{21}^{{(1)}}&= -262.1990984105955060153722678925879686999467800030\ldots,\\
M_{21}^{{(2)}}&=
2873.2506195273103871316630429729277291187753911438\ldots,\\
M_{21}^{{(3)}}&=
-20850.77252988625214437845006030730940289184143944\ldots,\\
M_{21}^{{(4)}}&=
123806.28243923947828962358530545969884163429504554\ldots,\\
M_{22}^{{(2)}}&=
1915.2828601394850124013600200661647497224250709008\ldots,\\
M_{22}^{{(3)}}&=
-9230.990439724705220633706650782234978383599409697\ldots,\\
M_{22}^{{(4)}}&=
24179.184087947556033170242057981283994228407821870\ldots,\\
M_{35}^{{(5)}}&= 20971.632987103828923594052378653898688768972653284\ldots,\\
M_{41}^{{(3)}}&=
30385.287535356282840631913693445372174956734072988\ldots,\\
M_{41}^{{(4)}}&=
-157029.840084536053310520481290249759485412480424\ldots.
\end{align}
The notation of these constants is as follows: $M_{A}^{(n)}$ denotes the expansion coefficient of $M_{A}(3-2\epsilon)/ M_{01}(3-2\epsilon)$ in front of $\epsilon^n$. 

We present results only for the integrals which can not be expressed in terms of $\Gamma$-functions. Where it is possible, we pull out rational factor making the expansion uniform in transcendental weight.
\begin{equation}
\frac{M_{21}(3-2\epsilon)}{M_{01}(3-2\epsilon)}=  M_{21}^{{(1)}}\epsilon+ M_{21}^{{(2)}}\epsilon ^2+ M_{21}^{{(3)}}\epsilon ^3+
M_{21}^{{(4)}}\epsilon ^4+O\left(\epsilon ^5\right)
\end{equation}
\begin{equation}
\frac{M_{22}(3-2\epsilon)}{M_{01}(3-2\epsilon)}=      M_{21}^{{(1)}}\epsilon+ M_{22}^{{(2)}}\epsilon ^2+ M_{22}^{{(3)}}\epsilon ^3+
M_{22}^{{(4)}}\epsilon ^4+O\left(\epsilon ^5\right)
\end{equation}
%\begin{multline}
%\frac{M_{23}(3-2\epsilon)}{M_{01}(3-2\epsilon)}= 
%-18 \zeta _2+  \left(204 \zeta _2-72 a_1 \zeta _2\right)\epsilon+ (-144 a_1^2 \zeta _2+816 a_1 \zeta _2\\-336 \zeta
%_2
%-900 \zeta _4)\epsilon ^2+ (-192 a_1^3 \zeta _2+1632 a_1^2 \zeta _2-1344 a_1 \zeta _2-3600 a_1 \zeta _4+384 \zeta
%_2\\
%-4356 \zeta _2 \zeta _3+10200 \zeta _4)\epsilon ^3+ (-192 a_1^4 \zeta _2+2176 a_1^3 \zeta _2-2688 a_1^2 \zeta _2-7200
%a_1^2 \zeta _4\\
%+1536 a_1 \zeta _2
%-17424 a_1 \zeta _2 \zeta _3+40800 a_1 \zeta _4+2304 \zeta _2+49368 \zeta _2 \zeta _3-16800 \zeta
%_4\\
%-{156681 \zeta _6}/{2})\epsilon ^4+O\left(\epsilon ^5\right)
%\end{multline}
\begin{multline}
\frac{M_{23}(3-2\epsilon)}{M_{01}(3-2\epsilon)}= 
-\frac{\pi ^2 2^{4 \epsilon } (4 \epsilon -1) (10 \epsilon -3) (10 \epsilon -1)}{6 \epsilon -1}
\bigg\{1+20 \zeta _2 \epsilon ^2+242 \zeta _3 \epsilon ^3\\
+2487 \zeta _4 \epsilon ^4+\left(4840 \zeta _2 \zeta _3+18054 \zeta _5\right) \epsilon ^5+\left(29282 \zeta _3^2+231135 \zeta _6\right) \epsilon ^6+O\left(\epsilon ^7\right)\bigg\}
\end{multline}
%\begin{multline}
%\frac{M_{24}(3-2\epsilon)}{M_{01}(3-2\epsilon)}= 
%-270 \zeta _4 \epsilon + \left(4680 \zeta _4-2160 a_1 \zeta _4\right)\epsilon ^2+ (-8640 a_1^2 \zeta _4+37440 a_1 \zeta
%_4\\
%-23400 \zeta _4-22680 \zeta _6)\epsilon ^3+ (-23040 a_1^3 \zeta _4+149760 a_1^2 \zeta _4-187200 a_1 \zeta _4-181440 a_1
%\zeta _6\\
%-76680 \zeta _3 \zeta _4+36000 \zeta _4+393120 \zeta _6)\epsilon ^4+O\left(\epsilon ^5\right)
%\end{multline}
\begin{multline}
\frac{M_{24}(3-2\epsilon)}{M_{01}(3-2\epsilon)}= 
\pi ^4 2^{8 \epsilon } \epsilon  (4 \epsilon -1) (10 \epsilon -3) (10 \epsilon -1)
\bigg\{
1+48 \zeta _2 \epsilon ^2+284 \zeta _3 \epsilon ^3+5706 \zeta _4 \epsilon ^4\\
+\left(13632 \zeta _2 \zeta _3+19356 \zeta _5\right) \epsilon ^5+\left(40328 \zeta _3^2+496040 \zeta _6\right) \epsilon ^6+O\left(\epsilon ^7\right)
\bigg\}
\end{multline}
%\begin{multline}
%\frac{M_{25}(3-2\epsilon)}{M_{01}(3-2\epsilon)}=
%9 \zeta _2+  \left(36 a_1 \zeta _2-210 \zeta _2\right)\epsilon+ (72 a_1^2 \zeta _2-840 a_1 \zeta _2+2040 \zeta _2\\
%-90\zeta _4)\epsilon ^2
%+ (96 a_1^3 \zeta _2-1680 a_1^2 \zeta _2+8160 a_1 \zeta _2-360 a_1 \zeta _4-13440 \zeta _2+666 \zeta
%_2 \zeta _3\\
%+2100 \zeta _4)\epsilon ^3+ (96 a_1^4 \zeta _2-2240 a_1^3 \zeta _2+16320 a_1^2 \zeta _2-720 a_1^2 \zeta
%_4-53760 a_1 \zeta _2+2664 a_1 \zeta _2 \zeta _3\\
% +8400 a_1 \zeta _4+80640 \zeta _2-15540 \zeta _2 \zeta _3-20400 \zeta _4-{13419
%	\zeta _6}/{4})\epsilon ^4+O\left(\epsilon ^5\right)
%\end{multline}
\begin{multline}
\frac{M_{25}(3-2\epsilon)}{M_{01}(3-2\epsilon)}=-\frac{\pi ^2 2^{4 \epsilon -1} (4 \epsilon -1) (10 \epsilon -3) (10 \epsilon -1)}{6 \epsilon +1}\bigg\{
1-4 \zeta _2 \epsilon ^2+74 \zeta _3 \epsilon ^3\\
-213 \zeta _4 \epsilon ^4+(1686 \zeta _5-296 \zeta _2 \zeta _3) \epsilon ^5+\left(2738 \zeta _3^2-5367 \zeta _6\right) \epsilon ^6+O\left(\epsilon ^7\right)
\bigg\}
\end{multline}
%\begin{multline}
%\frac{M_{26}(3-2\epsilon)}{M_{01}(3-2\epsilon)}=
%9 \zeta _2+  \left(-36 a_1 \zeta _2-210 \zeta _2+126 \zeta _3\right)\epsilon+ (-72 a_1^2 \zeta _2+840 a_1 \zeta _2+24
%a_1^4\\
%+576 a_4+2040 \zeta _2-2940 \zeta _3+243 \zeta _4)\epsilon ^2+ (96 a_1^3 \zeta _2+1680 a_1^2 \zeta _2-8160 a_1 \zeta
%_2-972 a_1 \zeta _4\\
%-{96 a_1^5}/{5}-560 a_1^4-13440 a_4+2304 a_5-13440 \zeta _2+126 \zeta _2 \zeta _3+28560 \zeta _3-5670 \zeta
%_4\\
%+3069 \zeta _5)\epsilon ^3+ (-96 a_1^4 \zeta _2-2240 a_1^3 \zeta _2-16320 a_1^2 \zeta _2+1944 a_1^2 \zeta _4+53760 a_1
%\zeta _2-504 a_1 \zeta _2 \zeta _3\\
%+22680 a_1 \zeta _4+{64 a_1^6}/{5}+448 a_1^5+5440 a_1^4+130560 a_4-53760 a_5+9216 a_6+3258
%\zeta _3^2\\
%+80640 \zeta _2-2940 \zeta _2 \zeta _3-188160 \zeta _3+55080 \zeta _4-71610 \zeta _5+{71901 \zeta _6}/{4}-6336
%s_6)\epsilon ^4\\
%+O\left(\epsilon ^5\right)
%\end{multline}
\begin{multline}
\frac{M_{26}(3-2\epsilon)}{M_{01}(3-2\epsilon)}=
-\frac{(4 \epsilon -1) (10 \epsilon -3) (10 \epsilon -1)}{6 \epsilon +1}
\bigg\{
3 \zeta _2+ \left(42 \zeta _3-12 a_1 \zeta _2\right)\epsilon \\
+ (-24 a_1^2 \zeta _2+8 a_1^4+192 a_4+81 \zeta _4)\epsilon ^2+ \bigg(32 a_1^3 \zeta _2-324 a_1 \zeta _4-\frac{32 a_1^5}{5}+768 a_5+42 \zeta _2 \zeta _3\\
+1023 \zeta _5\bigg)\epsilon ^3+ \bigg(-32 a_1^4 \zeta _2+648 a_1^2 \zeta _4-168 a_1 \zeta _2 \zeta _3+\frac{64 a_1^6}{15}+3072 a_6+1086 \zeta _3^2\\
+\frac{23967 \zeta _6}{4}-2112 s_6\bigg)\epsilon ^4+O\left(\epsilon ^5\right)
\bigg\}
\end{multline}
%\begin{multline}
%\frac{M_{27}(3-2\epsilon)}{M_{01}(3-2\epsilon)}=
%-18 \zeta _2 + \left(-144 a_1 \zeta _2+204 \zeta _2+126 \zeta _3\right)\epsilon+ (-288
%a_1^2 \zeta _2+1632 a_1 \zeta _2\\
%-48 a_1^4-1152 a_4-336 \zeta _2-1428 \zeta _3-891 \zeta
%_4)\epsilon ^2+ (-768 a_1^3 \zeta _2+3264 a_1^2 \zeta _2-2688 a_1 \zeta _2\\
%-7128 a_1
%\zeta _4-{384 a_1^5}/{5}+544 a_1^4+13056 a_4+9216 a_5+384 \zeta _2-5112 \zeta _2 \zeta _3+2352
%\zeta _3\\+10098 \zeta _4
%+3906 \zeta _5)\epsilon ^3 +(-1536 a_1^4 \zeta _2+8704 a_1^3
%\zeta _2-5376 a_1^2 \zeta _2-28512 a_1^2 \zeta _4+3072 a_1 \zeta _2\\
%-40896 a_1 \zeta _2 \zeta
%_3+80784 a_1 \zeta _4-{512 a_1^6}/{5}+{4352 a_1^5}/{5}-896 a_1^4-21504 a_4-104448 a_5\\
%-73728
%a_6+11844 \zeta _3^2+2304 \zeta _2+57936 \zeta _2 \zeta _3-2688 \zeta _3-16632 \zeta _4-44268 \zeta
%_5-56673 \zeta _6\\
%+16128 s_6)\epsilon ^4+O\left(\epsilon ^5\right)
%\end{multline}
\begin{multline}
\frac{M_{27}(3-2\epsilon)}{M_{01}(3-2\epsilon)}=
-\frac{(4 \epsilon -1) (10 \epsilon -3) (10 \epsilon -1)}{6 \epsilon -1}
\bigg\{
6 \zeta _2+  (48 a_1 \zeta _2
-42 \zeta _3)\epsilon+ (96 a_1^2 \zeta _2\\+16 a_1^4+384 a_4+297 \zeta _4)\epsilon ^2+ \bigg(256 a_1^3 \zeta _2+2376 a_1 \zeta _4+\frac{128 a_1^5}{5}-3072 a_5+1704 \zeta _2 \zeta _3\\
-1302 \zeta _5\bigg)\epsilon ^3+\bigg(512 a_1^4 \zeta _2+9504 a_1^2 \zeta _4+13632 a_1 \zeta _2 \zeta _3+\frac{512 a_1^6}{15}+24576 a_6-3948 \zeta _3^2\\
+18891 \zeta _6-5376 s_6\bigg)\epsilon ^4 +O\left(\epsilon ^5\right)
\bigg\}
\end{multline}
%\begin{multline}
%\frac{M_{32}(3-2\epsilon)}{M_{01}(3-2\epsilon)}=
%-270 \zeta _4 \epsilon +\left(4680 \zeta _4-1512 \zeta _2 \zeta _3\right) \epsilon ^2+
%(-288 a_1^4 \zeta _2+4320 a_1^2 \zeta _4\\
%-6048 a_1 \zeta _2 \zeta _3-6912 a_4 \zeta _2+26208
%\zeta _2 \zeta _3-23400 \zeta _4-22113 \zeta _6)\epsilon ^3+O\left(\epsilon ^4\right)
%\end{multline}
\begin{multline}
\frac{M_{32}(3-2\epsilon)}{M_{01}(3-2\epsilon)}=
\pi ^2 (4 \epsilon -1) (10 \epsilon -3) (10 \epsilon -1)\epsilon
\bigg\{
6 \zeta _2+84 \zeta _3 \epsilon + (-96 a_1^2 \zeta _2\\
+336 a_1 \zeta _3+16 a_1^4+384 a_4+702 \zeta _4)\epsilon ^2+ \bigg(-256 a_1^3 \zeta _2+672 a_1^2 \zeta _3+\frac{256 a_1^5}{5}+1536 a_4 a_1\\
+1536 a_5+4464 \zeta _2 \zeta _3+2046 \zeta _5\bigg)\epsilon ^3+ \bigg(192 a_1^4 \zeta _2+896 a_1^3 \zeta _3-8640 a_1^2 \zeta _4+12096 a_1 \zeta _2 \zeta _3\\+8184 a_1 \zeta _5
+13824 a_4 \zeta _2+\frac{256 a_1^6}{3}+3072 a_4 a_1^2+6144 a_5 a_1+6144 a_6+21156 \zeta _3^2+65691 \zeta _6\\
-4224 s_6\bigg)\epsilon ^4+O\left(\epsilon ^5\right)
\bigg\}
\end{multline}
%\begin{multline}
%\frac{M_{33}(3-2\epsilon)}{M_{01}(3-2\epsilon)}=
%18 \zeta _2+  \left(72 a_1 \zeta _2-276 \zeta _2\right)\epsilon+ (144 a_1^2 \zeta
%_2-1104 a_1 \zeta _2-144 \zeta _2\\
%+2700 \zeta _4)\epsilon ^2+ (192 a_1^3 \zeta _2-2208
%a_1^2 \zeta _2-576 a_1 \zeta _2+28080 a_1 \zeta _4+31104 \zeta _2-5004 \zeta _2 \zeta _3\\
%-58680 \zeta
%_4)\epsilon ^3+ (2496 a_1^4 \zeta _2-2944 a_1^3 \zeta _2-1152 a_1^2 \zeta _2+125280 a_1^2
%\zeta _4+124416 a_1 \zeta _2\\
%-20016 a_1 \zeta _2 \zeta _3-637920 a_1 \zeta _4+55296 a_4 \zeta
%_2-401664 \zeta _2+125112 \zeta _2 \zeta _3+485280 \zeta _4\\
%+{270837 \zeta_6}/{2})\epsilon ^4+O\left(\epsilon ^5\right)
%\end{multline}
\begin{multline}
\frac{M_{33}(3-2\epsilon)}{M_{01}(3-2\epsilon)}=
-\frac{\pi ^2 (4 \epsilon -1) (10 \epsilon -3) (10 \epsilon -1)}{(6 \epsilon +1)^2}\bigg\{
1+\left(4 a_1+14\right) \epsilon + (8 a_1^2+56 a_1\\
+60 \zeta _2)\epsilon ^2+ \left(624 a_1 \zeta _2+\frac{32 a_1^3}{3}+112 a_1^2+456 \zeta _2-278 \zeta _3\right)\epsilon ^3+ \bigg(2784 a_1^2 \zeta _2+4128 a_1 \zeta _2\\
-1112 a_1 \zeta _3+\frac{416 a_1^4}{3}+\frac{448 a_1^3}{3}+3072 a_4-1204 \zeta _3+4299 \zeta _4\bigg)\epsilon ^4+ \bigg(8832 a_1^3 \zeta _2\\
+17472 a_1^2 \zeta _2-2224 a_1^2 \zeta _3-4816 a_1 \zeta _3+47724 a_1 \zeta _4+\frac{2176 a_1^5}{3}+\frac{2752 a_1^4}{3}+12288 a_4 a_1\\
+18432 a_4-24576 a_5+13080 \zeta _2 \zeta _3+29658 \zeta _4-8922 \zeta _5\bigg)\epsilon ^5+\bigg(24704 a_1^4 \zeta _2+54016 a_1^3 \zeta _2\\
-\frac{8896}{3} a_1^3 \zeta _3-9632 a_1^2 \zeta _3+255576 a_1^2 \zeta _4+183648 a_1 \zeta _2 \zeta _3+301800 a_1 \zeta _4-35688 a_1 \zeta _5\\
+36864 a_4 \zeta _2+\frac{95488 a_1^6}{45}+\frac{66304 a_1^5}{15}+24576 a_4 a_1^2+73728 a_4 a_1-98304 a_5 a_1-147456 a_5\\
+196608 a_6-49390 \zeta _3^2+84048 \zeta _2 \zeta _3-41580 \zeta _5+256873 \zeta _6-43008 s_6\bigg)\epsilon ^6 +O\left(\epsilon ^7\right)
\bigg\}
\end{multline}
%\begin{multline}
%\frac{M_{34}(3-2\epsilon)}{M_{01}(3-2\epsilon)}=
%9 \zeta _2+  \left(36 a_1 \zeta _2-210 \zeta _2\right)\epsilon+ (72 a_1^2 \zeta _2-840
%a_1 \zeta _2+2040 \zeta _2\\
%-1170 \zeta _4)\epsilon ^2
%+ (96 a_1^3 \zeta _2-1680 a_1^2 \zeta
%_2+8160 a_1 \zeta _2-21960 a_1 \zeta _4-13440 \zeta _2-630 \zeta _2 \zeta _3\\+27300 \zeta
%_4)\epsilon ^3
%+ (2400 a_1^4 \zeta _2-2240 a_1^3 \zeta _2+16320 a_1^2 \zeta _2-113040 a_1^2\zeta _4-53760 a_1 \zeta _2\\-2520 a_1 \zeta _2 \zeta _3
%+512400 a_1 \zeta _4+55296 a_4 \zeta _2+80640
%\zeta _2+14700 \zeta _2 \zeta _3-265200 \zeta _4\\
%-{796131 \zeta _6}/{4})\epsilon ^4+O\left(\epsilon^5\right)
%\end{multline}
\begin{multline}
\frac{M_{34}(3-2\epsilon)}{M_{01}(3-2\epsilon)}=
-\frac{\pi ^2 (4 \epsilon -1) (10 \epsilon -3) (10 \epsilon -1)}{2 (6 \epsilon +1)}\bigg\{
1+4 a_1 \epsilon + \left(8 a_1^2-52 \zeta _2\right)\epsilon ^2\\
+ \left(-976 a_1 \zeta _2+\frac{32 a_1^3}{3}-70 \zeta _3\right)\epsilon ^3+ \bigg(-5024 a_1^2 \zeta _2-280 a_1 \zeta _3+\frac{800 a_1^4}{3}+6144 a_4\\
-12637 \zeta _4\bigg)\epsilon ^4+ \bigg(-\frac{38528}{3} a_1^3 \zeta _2-560 a_1^2 \zeta _3-167668 a_1 \zeta _4+\frac{15488 a_1^5}{15}+24576 a_4 a_1\\
-13128 \zeta _2 \zeta _3-12234 \zeta _5\bigg)\epsilon ^5+ \bigg(-\frac{60032}{3} a_1^4 \zeta _2-\frac{2240}{3} a_1^3 \zeta _3-584936 a_1^2 \zeta _4-269088 a_1 \zeta _2 \zeta _3\\
-48936 a_1 \zeta _5+24576 a_4 \zeta _2+\frac{92416 a_1^6}{45}+49152 a_4 a_1^2+101458 \zeta _3^2-903075 \zeta _6\\
-270336 s_6\bigg)\epsilon ^6+O\left(\epsilon ^7\right)
\bigg\}
\end{multline}
%\begin{multline}
%\frac{M_{35}(3-2\epsilon)}{M_{01}(3-2\epsilon)}=
%-\frac{3}{\epsilon}+52+\left(-120\zeta_2-80\right)\epsilon+\left(-864a_1\zeta_2+2944\zeta_2+1056\zeta_3-5024\right)\epsilon^2\\
%+(-1728a_1^2\zeta_2+20736a_1\zeta_2-192a_1^4-4608a_4-30272\zeta_2-22336\zeta_3-16272\zeta_4\\
%+76992)\epsilon^3
%+(-256M_{21}^{{(1)}}-3840a_1^3\zeta_2+41472a_1^2\zeta_2-203520a_1\zeta_2-148896
%a_1\zeta_4-{768a_1^5}/{5}\\
%+4096a_1^4+98304a_4+18432a_5+201088\zeta_2-54192\zeta_2
%\zeta_3+154496\zeta_3+396384\zeta_4+101016\zeta_5\\
%-770688)\epsilon^4+
%M_{35}^{{(5)}}\epsilon^5+O\left(\epsilon^6\right)
%\end{multline}
\begin{multline}
\frac{M_{35}(3-2\epsilon)}{M_{01}(3-2\epsilon)}=\frac{(4 \epsilon -1) (10 \epsilon -3) (10 \epsilon -1)}{  (2 \epsilon +1) (6 \epsilon +1)^2}
\bigg\{
\frac1\epsilon+14 +40 \zeta _2 \epsilon+ (288 a_1 \zeta _2\\
+272 \zeta _2-352 \zeta _3)\epsilon ^2+ \left(576 a_1^2 \zeta _2+2112 a_1 \zeta _2+64 a_1^4+1536 a_4-3584 \zeta _3+5424 \zeta _4\right)\epsilon ^3\\
+ \bigg(\frac{256 M_{21}^{(1)}}{3}+1280 a_1^3 \zeta _2+4224 a_1^2 \zeta _2+49632 a_1 \zeta _4+\frac{256 a_1^5}{5}+640 a_1^4+15360 a_4\\
-6144 a_5+18064 \zeta _2 \zeta _3+37824 \zeta _4-33672 \zeta _5\bigg)\epsilon ^4+ \bigg(\frac{24064
	M_{21}^{\text{(1)}}}{9}-\frac{M_{35}^{\text{(5)}}}{3}+\frac{120320}{3} a_1^3 \zeta _2\\
-135680 a_1^2
\zeta _2+411136 a_1 \zeta _2+1555136 a_1 \zeta _4+\frac{24064 a_1^5}{15}-9728 a_1^4-233472 a_4\\
-192512
a_5-367360 \zeta _2+\frac{1698016 \zeta _2 \zeta _3}{3}-\frac{107776 \zeta _3}{3}-1338816 \zeta
_4-1055056 \zeta _5\\
+2177280\bigg)\epsilon^5+O\left(\epsilon ^6\right)
\bigg\}
\end{multline}
%\begin{multline}
%\frac{M_{36}(3-2\epsilon)}{M_{01}(3-2\epsilon)}=
%-\frac{6}{\epsilon }+92-\left(144 \zeta _2+120\right) \epsilon + \left(-576 a_1 \zeta
%_2+3168 \zeta _2-576 \zeta _3-4432\right)\epsilon ^2\\
%+ \left(12288 a_1 \zeta
%_2-1152 a_1^2 \zeta _2+192 a_1^4+4608 a_4-24512 \zeta _2+12192 \zeta _3-16584 \zeta _4+33312\right)\epsilon ^3\\
%+
%(24576 a_1^2 \zeta _2-87552 a_1 \zeta _2-47520 a_1 \zeta _4-3328 a_1^4-79872 a_4+58752 \zeta
%_2-27744 \zeta _2 \zeta _3\\
%-94144 \zeta _3+325904 \zeta _4-73512 \zeta _5-62784)\epsilon ^4+
%(-2304 a_1^4 \zeta _2+6144 a_1^3 \zeta _2-175104 a_1^2 \zeta _2\\
%+34560 a_1^2 \zeta _4+107520 a_1
%\zeta _2-89856 a_1 \zeta _2 \zeta _3+892800 a_1 \zeta _4-55296 a_4 \zeta _2+11264 a_1^4+270336
%a_4\\
%-39984 \zeta _3^2+647424 \zeta _2+531968 \zeta _2 \zeta _3+382848 \zeta _3-1954720 \zeta
%_4+1472928 \zeta _5-1285912 \zeta _6\\
%-202752 s_6-1058688)\epsilon ^5+O\left(\epsilon ^6\right)
%\end{multline}
\begin{multline}
\frac{M_{36}(3-2\epsilon)}{M_{01}(3-2\epsilon)}=
-\frac{(4 \epsilon -1) (10 \epsilon -3) (10 \epsilon -1)}{(2 \epsilon +1)^2 (6 \epsilon +1)^2}
\bigg\{
-\frac{2}{\epsilon }-36+\left(-48 \zeta _2-176\right) \epsilon \\
+ \left(-192 a_1 \zeta _2-544 \zeta _2-192 \zeta _3\right)\epsilon ^2+(-384 a_1^2 \zeta _2-2304 a_1 \zeta _2+64 a_1^4+1536 a_4\\
-768 \zeta _2-2336 \zeta _3-5528 \zeta _4)\epsilon ^3 + (-4608 a_1^2 \zeta _2-3840 a_1 \zeta _2-15840 a_1 \zeta _4+1024 a_1^4\\
+24576 a_4-9248 \zeta _2 \zeta _3-7104 \zeta _3-75632 \zeta _4-24504 \zeta _5)\epsilon ^4+\bigg(-768 a_1^4 \zeta _2+2048 a_1^3 \zeta _2\\
-7680 a_1^2 \zeta _2+11520 a_1^2 \zeta _4-29952 a_1 \zeta _2 \zeta _3-230400 a_1 \zeta _4-18432 a_4 \zeta _2+3840 a_1^4+92160 a_4\\
-13328 \zeta _3^2-130944 \zeta _2 \zeta _3-231744 \zeta _4-325824 \zeta _5-\frac{1285912 \zeta _6}{3}-67584 s_6\bigg)\epsilon ^5 \\
+O\left(\epsilon ^6\right)
\bigg\}
\end{multline}
\begin{multline}
\frac{M_{41}(3-2\epsilon)}{M_{01}(3-2\epsilon)}=-\frac{3}{\epsilon}+9\zeta_2+40+\left({3M_{21}^{{(1)}}}/{2}-180
a_1\zeta_2+162\zeta_2+378\zeta_3-88\right)\epsilon\\
+(14M_{21}^{{(1)}}+{3
	M_{22}^{{(2)}}}/{2}-360a_1^2\zeta_2+2472a_1\zeta_2+72a_1^4+1728a_4-5992\zeta_2-1848
\zeta_3+3069\zeta_4\\
+1616)\epsilon^2
+M_{41}^{{(3)}}\epsilon^3+M_{41}^{{(4)}}\epsilon^4+O\left(\epsilon^5\right)
\end{multline}
\begin{multline}
\frac{M_{42}(3-2\epsilon)}{M_{01}(3-2\epsilon)}=
-\frac{3}{\epsilon }+40+\left(396 \zeta _2-405 \zeta _4+368\right) \epsilon + (-24
M_{21}^{{(1)}}+1728 a_1 \zeta _2\\
+1620 a_1 \zeta _4-4632 \zeta _2-1638 \zeta _2 \zeta _3+3204
\zeta _3+3240 \zeta _4-2790 \zeta _5-12432)\epsilon ^2+ (-120 M_{21}^{{(1)}}\\
-12
M_{21}^{{(2)}}-24 M_{22}^{{(2)}}-{3 M_{35}^{{(5)}}}/{64}-420 a_1^4 \zeta _2+4128
a_1^3 \zeta _2-15624 a_1^2 \zeta _2+38520 a_1 \zeta _2\\
-4032 a_4 \zeta _2-17631 a_1 \zeta _2 \zeta
_3-22779 a_1^2 \zeta _4+145656 a_1 \zeta _4-\frac{24 a_1^6}{5}+\frac{768 a_1^5}{5}-1368 a_1^4\\-32832
a_4
-18432 a_5-3456 a_6+{6777 \zeta _3^2}/{4}-77436 \zeta _2+73254 \zeta _2 \zeta _3-77436 \zeta
_3\\-149373 \zeta _4
-78588 \zeta _5
-{3933495 \zeta _6}/{32}+15480 s_6+459588)\epsilon ^3+O\left(\epsilon
^4\right)
\end{multline}
\begin{multline}
\frac{M_{43}(3-2\epsilon)}{M_{01}(3-2\epsilon)}=
\frac{48 \zeta _2}{5}-\frac{156}{5}+\left(\frac{288 a_1 \zeta _2}{5}-\frac{3588
	\zeta _2}{25}+\frac{408 \zeta _3}{5}+\frac{5976}{25}\right)\epsilon  + \bigg(\frac{576}{5} a_1^2
\zeta _2\\
-\frac{12528 a_1 \zeta _2}{25}-\frac{192 a_1^4}{5}-\frac{4608 a_4}{5}-\frac{416416 \zeta
	_2}{375}-\frac{35448 \zeta _3}{25}+\frac{5916 \zeta _4}{5}+\frac{1401232}{375}\bigg)\epsilon ^2\\
+
\bigg(-\frac{768}{5} a_1^3 \zeta _2-\frac{25056}{25} a_1^2 \zeta _2-\frac{1192832 a_1 \zeta
	_2}{125}+\frac{7776 a_1 \zeta _4}{5}+\frac{768 a_1^5}{25}+\frac{8352 a_1^4}{25}+\frac{200448
	a_4}{25}\\
-\frac{18432 a_5}{5}+\frac{51526496 \zeta _2}{1875}-816 \zeta _2 \zeta _3+\frac{740864 \zeta
	_3}{375}-\frac{355596 \zeta _4}{25}+\frac{39288 \zeta _5}{5}-\frac{26162464}{625}\bigg)\epsilon ^3\\
+
\bigg(\frac{33408}{25} a_1^3 \zeta _2-\frac{2385664}{125} a_1^2 \zeta _2-\frac{4032}{5} a_1^2 \zeta
_4+\frac{78916992 a_1 \zeta _2}{625}-\frac{12096}{5} a_1 \zeta _2 \zeta _3-\frac{338256 a_1 \zeta
	_4}{25}\\
-\frac{18432 a_4 \zeta _2}{5}-\frac{512 a_1^6}{25}-\frac{33408 a_1^5}{125}+\frac{1003264
	a_1^4}{375}+\frac{8026112 a_4}{125}+\frac{801792 a_5}{25}-\frac{73728 a_6}{5}\\
-\frac{15792 \zeta
	_3^2}{5}-\frac{988918976 \zeta _2}{9375}+\frac{43176 \zeta _2 \zeta _3}{5}+\frac{184730816 \zeta
	_3}{1875}-\frac{28598272 \zeta _4}{375}-\frac{2507028 \zeta _5}{25}\\
+\frac{217402 \zeta
	_6}{5}+\frac{50688 s_6}{5}+\frac{849324352}{9375}\bigg)\epsilon ^4+O\left(\epsilon ^5\right)
\end{multline}
\begin{multline}
\frac{M_{44}(3-2\epsilon)}{M_{01}(3-2\epsilon)}=
-\frac{6}{\epsilon }+92+\left(-192 \zeta _2-120\right) \epsilon + (3712 \zeta _2-576 a_1 \zeta
_2-1104 \zeta _3\\
-4432)\epsilon ^2
+(16 M_{21}^{{(1)}}-1152 a_1^2 \zeta
_2+12288 a_1 \zeta _2+96 a_1^4+2304 a_4-21760 \zeta _2+14720 \zeta _3\\
-17352 \zeta_4+33312)\epsilon ^3 + (576 M_{21}^{{(1)}}+16 M_{22}^{{(2)}}-768 a_1^3 \zeta
_2+24576 a_1^2 \zeta _2-96768 a_1 \zeta _2\\-36432 a_1 \zeta _4
+{384 a_1^5}/{5}-512 a_1^4-12288
a_4-9216 a_5+10752 \zeta _2-21504 \zeta _2 \zeta _3+55936 \zeta _3\\+326976 \zeta _4
-100476 \zeta
_5-62784)\epsilon ^4+O\left(\epsilon ^5\right)
\end{multline}
\begin{multline}
\frac{M_{45}(3-2\epsilon)}{M_{01}(3-2\epsilon)}=
-\frac{6}{\epsilon }+104+\left(224-432 \zeta _2\right) \epsilon + (-5184 a_1 \zeta
_2+10944 \zeta _2+8352 \zeta _3\\
-22848)\epsilon ^2+ (-10368 a_1^2 \zeta _2+112896 a_1 \zeta
_2-2304 a_1^4-55296 a_4-121344 \zeta _2-151680 \zeta _3\\
-56016 \zeta _4+358272)\epsilon ^3+
(-2048 M_{21}^{{(1)}}-32256 a_1^3 \zeta _2+225792 a_1^2 \zeta _2-933888 a_1 \zeta _2\\
-875520
a_1 \zeta _4-3072 a_1^5+43008 a_1^4+1032192 a_4+368640 a_5+935936 \zeta _2-372384 \zeta _2 \zeta
_3\\
+643584 \zeta _3+1461696 \zeta _4+887328 \zeta _5-3808512)\epsilon ^4+ \bigg(-13312
M_{21}^{{(1)}}+512 M_{21}^{{(2)}}\\
+10 M_{35}^{{(5)}}+27648 a_1^4 \zeta _2-235520 a_1^3
\zeta _2+2202624 a_1^2 \zeta _2-7836672 a_1 \zeta _2+884736 a_4 \zeta _2\\
+921600 a_1 \zeta _2 \zeta
_3+10368 a_1^2 \zeta _4-15579648 a_1 \zeta _4-\frac{13312 a_1^6}{5}+\frac{126976 a_1^5}{5}+78848
a_1^4\\
+1892352 a_4-3047424 a_5-1916928 a_6-846240 \zeta _3^2+4563456 \zeta _2
-4577920 \zeta _2 \zeta
_3\\+5038592 \zeta _3+23521920 \zeta _4+4482048 \zeta _5+9846480 \zeta _6+1483776
s_6-31555584\bigg)\epsilon ^5\\
+O\left(\epsilon ^6\right)
\end{multline}
\begin{multline}
\frac{M_{51}(3-2\epsilon)}{M_{01}(3-2\epsilon)}=
-\frac{6}{\epsilon }+\left(\frac{128 \zeta _2}{5}-\frac{6}{5}\right)+  \left(768 a_1 \zeta
_2-\frac{49744 \zeta _2}{75}-448 \zeta _3+\frac{78428}{75}\right)\epsilon\\
+\bigg(1536 a_1^2 \zeta
_2-\frac{72896 a_1 \zeta _2}{5}+\frac{3584 a_1^4}{5}+\frac{86016 a_4}{5}+\frac{442688 \zeta
	_2}{125}+\frac{18880 \zeta _3}{3}-9568 \zeta _4\\
+\frac{624232}{375}\bigg)\epsilon ^2 +
\bigg(\frac{38912}{5} a_1^3 \zeta _2-\frac{145792}{5} a_1^2 \zeta _2+\frac{1792256 a_1 \zeta
	_2}{25}-28416 a_1 \zeta _4+\frac{8192 a_1^5}{5}\\
-\frac{958912 a_1^4}{75}-\frac{7671296 a_4}{25}-196608
a_5+\frac{21599872 \zeta _2}{625}+\frac{37248 \zeta _2 \zeta _3}{5}+5120 \zeta _3+\frac{2726024 \zeta
	_4}{15}\\
+85056 \zeta _5-\frac{282285392}{1875}\bigg)\epsilon ^3+ \bigg(\frac{57344
	M_{21}^{{(1)}}}{15}-\frac{4096 M_{21}^{{(2)}}}{15}-\frac{16
	M_{35}^{{(5)}}}{15}+\frac{90112}{5} a_1^4 \zeta _2\\
-\frac{1180672}{25} a_1^3 \zeta
_2-\frac{7269888}{25} a_1^2 \zeta _2+\frac{183588864 a_1 \zeta _2}{125}-\frac{49152 a_4 \zeta
	_2}{5}-\frac{428544}{5} a_1 \zeta _2 \zeta _3\\
-\frac{1314816}{5} a_1^2 \zeta _4+\frac{22281184 a_1
	\zeta _4}{5}+\frac{57344 a_1^6}{25}-\frac{667648 a_1^5}{25}+\frac{13065472
	a_1^4}{375}+\frac{104523776 a_4}{125}\\
+\frac{16023552 a_5}{5}+\frac{8257536 a_6}{5}+\frac{453248 \zeta
	_3^2}{5}-\frac{15643997696 \zeta _2}{9375}+\frac{93949856 \zeta _2 \zeta _3}{75}-\frac{7967104 \zeta
	_3}{15}\\
-\frac{141720768 \zeta _4}{25}-\frac{20465288 \zeta _5}{5}-\frac{43744336 \zeta_6}{15}-\frac{172032 s_6}{5}+\frac{80950072352}{9375}\bigg)\epsilon ^4+O\left(\epsilon ^5\right)
\end{multline}
\begin{multline}
\frac{M_{61}(3-2\epsilon)}{M_{01}(3-2\epsilon)}=
\frac{108}{5 \epsilon }-\frac{224 \zeta _2}{5}-\frac{5564}{25}+ 
\bigg(\frac{168596 \zeta _2}{75}-\frac{1344}{5} a_1 \zeta _2-\frac{112 \zeta_3}{5}\\
-\frac{134296}{75}\bigg)\epsilon
+ \bigg(-\frac{2688}{5} a_1^2 \zeta _2+\frac{279952 a_1
	\zeta _2}{25}+\frac{448 a_1^4}{5}+\frac{10752 a_4}{5}-\frac{3532096 \zeta _2}{225}\\
-\frac{514472 \zeta_3}{75}-\frac{32312 \zeta _4}{5}
+\frac{59912272}{1875}\bigg)\epsilon ^2+\bigg(\frac{15712
	M_{21}^{{(1)}}}{225}+\frac{559904}{25} a_1^2 \zeta _2-\frac{5176832 a_1 \zeta _2}{75}\\
-30240 a_1\zeta _4
+\frac{53024 a_1^4}{25}+\frac{1272576 a_4}{25}-\frac{792795808 \zeta _2}{5625}+\frac{58912
	\zeta _2 \zeta _3}{5}+\frac{132903872 \zeta _3}{1125}\\
+\frac{19776236 \zeta _4}{75}
-45304 \zeta
_5-\frac{6109675552}{28125}\bigg)\epsilon ^3 + \bigg(\frac{13531904
	M_{21}^{{(1)}}}{3375}-\frac{1792 M_{21}^{{(2)}}}{15}+\frac{15712
	M_{22}^{{(2)}}}{225}\\-\frac{7 M_{35}^{{(5)}}}{15}
+\frac{60928}{15} a_1^4 \zeta
_2+\frac{2198144}{25} a_1^3 \zeta _2-\frac{24600064}{75} a_1^2 \zeta _2-\frac{178434176 a_1 \zeta
	_2}{1875}+157696 a_4 \zeta _2\\
-\frac{237888}{5} a_1 \zeta _2 \zeta _3-384832 a_1^2 \zeta
_4+\frac{17562416 a_1 \zeta _4}{5}+\frac{35456 a_1^5}{15}-\frac{60889856 a_1^4}{1125}-\frac{487118848
	a_4}{375}\\
-283648 a_5+80640 \zeta _3^2+\frac{31107200576 \zeta _2}{28125}+\frac{24326024 \zeta _2
	\zeta _3}{25}-\frac{182462144 \zeta _3}{375}\\-\frac{4725215776 \zeta _4}{1125}
-\frac{35436356 \zeta
	_5}{25}-\frac{4075022 \zeta _6}{3}+\frac{39424
	s_6}{5}+\frac{1995434100544}{421875}\bigg)\epsilon ^4\\
+O\left(\epsilon ^5\right)
\end{multline}
\begin{multline}
\frac{M_{62}(3-2\epsilon)}{M_{01}(3-2\epsilon)}=
\frac{192}{5 \epsilon }+\frac{1344 \zeta _2}{5}-\frac{28896}{25}+
\bigg(\frac{165232 \zeta _2}{25}-\frac{56448}{5} a_1 \zeta _2+\frac{81312 \zeta_3}{5}\\
-\frac{362192}{25}\bigg)\epsilon
+ \bigg(\frac{392 M_{21}^{{(1)}}}{5}-\frac{112896}{5}
a_1^2 \zeta _2+295104 a_1 \zeta _2-\frac{91392 a_1^4}{5}
-\frac{2193408 a_4}{5}\\
-\frac{111266224 \zeta_2}{375}
-\frac{10550208 \zeta _3}{25}+\frac{1340304 \zeta_4}{5}+\frac{1237871728}{1875}\bigg)\epsilon ^2+ \bigg(-\frac{114352
	M_{21}^{{(1)}}}{75}\\
+\frac{392 M_{22}^{{(2)}}}{5}-\frac{881664}{5} a_1^3 \zeta _2
+590208
a_1^2 \zeta _2-\frac{413288512 a_1 \zeta _2}{125}+\frac{2346624 a_1 \zeta _4}{5}-\frac{795648
	a_1^5}{25}\\
+\frac{7470464 a_1^4}{25}+\frac{179291136 a_4}{25}
+\frac{19095552 a_5}{5}+\frac{18417516896
	\zeta _2}{5625}-581952 \zeta _2 \zeta _3\\
+\frac{1384628608 \zeta _3}{375}-\frac{66637664 \zeta
	_4}{25}-\frac{3297504 \zeta _5}{5}
-\frac{722104920704}{84375}\bigg)\epsilon ^3+
\bigg(\frac{43008 M_{21}^{{(2)}}}{5}\\
-\frac{28827616 M_{21}^{{(1)}}}{375}-\frac{169232
	M_{22}^{{(2)}}}{75}+\frac{784 M_{41}^{{(3)}}}{15}+\frac{168
	M_{35}^{{(5)}}}{5}
-\frac{1562624}{5} a_1^4 \zeta _2\\
+\frac{5338112}{25} a_1^3 \zeta
_2+\frac{850690176}{125} a_1^2 \zeta _2-\frac{39064847488 a_1 \zeta _2}{1875}+\frac{172032 a_4 \zeta
	_2}{5}\\
+\frac{18982656}{5} a_1 \zeta _2 \zeta _3+9606912 a_1^2 \zeta _4-\frac{2701956096 a_1 \zeta
	_4}{25}-\frac{243712 a_1^6}{5}+\frac{53266432 a_1^5}{125}\\-\frac{19590528 a_1^4}{125}
-\frac{470172672
	a_4}{125}-\frac{1278394368 a_5}{25}-35094528 a_6-1659840 \zeta _3^2\\
+\frac{37182753856 \zeta_2}{1875}-\frac{729078848 \zeta _2 \zeta _3}{25}
-\frac{2029091008 \zeta _3}{1875}+\frac{13334743136
	\zeta _4}{125}\\
+\frac{1680498624 \zeta _5}{25}+\frac{380409064 \zeta _6}{5}+\frac{5892096
	s_6}{5}
-\frac{179237261924672}{1265625}\bigg)\epsilon ^4+O\left(\epsilon ^5\right)
\end{multline}
\begin{multline}
\frac{M_{63}(3-2\epsilon)}{M_{01}(3-2\epsilon)}=
-\frac{30}{\epsilon}+48\zeta_2+\frac{22}{5}+\left(9504a_1\zeta_2-998
\zeta_2-6888\zeta_3+\frac{48788}{25}\right)\epsilon\\
+\bigg(28M_{21}^{{(1)}}+19008a_1^2
\zeta_2-109032a_1\zeta_2+11424a_1^4+274176a_4-\frac{15192\zeta_2}{5}+\frac{125692\zeta
	_3}{5}\\
-141588\zeta_4+\frac{105856648}{1125}\bigg)\epsilon^2+\bigg(\frac{10696
	M_{21}^{{(1)}}}{5}+28M_{22}^{{(2)}}+116736a_1^3\zeta_2-218064a_1^2\zeta
_2\\
-\frac{859552a_1\zeta_2}{5}-814320a_1\zeta_4+24576a_1^5-\frac{702736
	a_1^4}{5}-\frac{16865664a_4}{5}-2949120a_5\\
+\frac{51276928\zeta_2}{225}+463536\zeta_2\zeta
_3+\frac{18373168\zeta_3}{25}+\frac{14592814\zeta_4}{5}+1144380\zeta
_5\\-\frac{10884323888}{5625}\bigg)\epsilon^3
+\bigg(\frac{7617712M_{21}^{{(1)}}}{75}-4096
M_{21}^{{(2)}}+\frac{28168M_{22}^{{(2)}}}{15}
+\frac{56M_{41}^{{(3)}}}{3}\\
-16
M_{35}^{{(5)}}+344704a_1^4\zeta_2
-\frac{39488}{5}a_1^3\zeta_2-\frac{34743744}{5}a_1^2
\zeta_2+\frac{578604032a_1\zeta_2}{25}\\
+1637376a_4\zeta_2+12672a_1\zeta_2\zeta_3
-5156736
a_1^2\zeta_4+\frac{394249896a_1\zeta_4}{5}+\frac{173824a_1^6}{5}\\
-\frac{6743232
	a_1^5}{25}-\frac{12002624a_1^4}{25}
-\frac{288062976a_4}{25}+\frac{161837568a_5}{5}+25030656
a_6\\
+1519512\zeta_3^2
-\frac{7949697856\zeta_2}{375}
+20969628\zeta_2\zeta_3
+\frac{1712657728
	\zeta_3}{1125}-\frac{6405160952\zeta_4}{75}\\
-65011022\zeta_5
-41811552\zeta_6
-2042880
s_6+\frac{3579868183328}{28125}\bigg)\epsilon^4+O\left(\epsilon^5\right)
\end{multline}

%\bibliography{SummerTime}

\end{document}